\documentclass{aa}

\usepackage{txfonts}
\usepackage{graphicx}

\begin{document}

   \title{Wide-field multi-color photometry of the Galactic globular cluster NGC
   1261.
   \thanks{Based on observations with the 1.3 m Warsaw telescope at Las Campanas
Observatory}}

      \author{V. Kravtsov\inst{1,2}
          \and
              G. Alca\'ino\inst{3}
          \and
              G. Marconi\inst{4}
           \and
              F. Alvarado\inst{3}
              }

\offprints{V. Kravtsov}

   \institute{Instituto de Astronom\'ia, Universidad Cat\'olica del Norte,
              Avenida Angamos 0610, Antofagasta, Chile\\
              \email{vkravtsov@ucn.cl}
            \and
              Sternberg Astronomical Institute, University Avenue 13,
              119899 Moscow, Russia\\
            \and
              Isaac Newton Institute of Chile, Ministerio de Educaci\'on de Chile,
              Casilla 8-9, Correo 9, Santiago, Chile\\
              \email{inewton@terra.cl, falvarad@eso.org}
            \and
              ESO - European Southern Observatory, Alonso de Cordova 3107, Vitacura,
              Santiago, Chile\\
              \email{gmarconi@eso.org}
             }

   \date{Received xxxxx / Accepted xxxxx}

   \abstract
{} {This work studies in more detail the stellar population,
including its photometric properties and characteristics, in the
rarely studied southern Galactic globular cluster NGC 1261. We focus
on the brighter sequences of the cluster's color-magnitude diagram
(CMD). Like in our previous works, we rely upon photometry in
several passbands to achieve more reliable results and conclusions.}
{We carried out and analyzed new multi-color photometry of NGC 1261
in $UBVI$ reaching below the turnoff point in all passbands in a
fairly extended cluster field, about 14$\arcmin$x14$\arcmin$.} {We found
several signs of the inhomogeneity ("multiplicity") in the stellar
population. The most prominent of them are: (1) the
dependence of the radial distribution of sub-giant branch (SGB)
stars in the cluster on their $U$ magnitude, with brighter stars
less centrally concentrated at the 99.9 \% level than their fainter
counterparts; (2) the dependence of the location of red giant branch
(RGB) stars in the $U$-$(U-B)$ CMD on their radial distance from the
cluster center, with the portion of stars bluer in the $(U-B)$ color
increasing towards the cluster outskirts. Additionally, the radial
variation of the RGB luminosity function in the bump region is
suspected. We assume that both the SGB stars brighter in the $U$ and
the RGB stars bluer in the $(U-B)$ color are probably associated
with blue horizontal branch stars, because of a
similarity in their radial distribution in the cluster. We estimated
the metalicity of NGC 1261 from the slope of the RGB in $U$-based CMDs
and the location of the RGB bump on the branch. These metallicity
indicators give [Fe/H]$_{ZW} = -1.34 \pm 0.16$ dex and [Fe/H]$_{ZW}
= -1.41 \pm 0.10$ dex, respectively. We isolated 18 probable blue
straggler candidates. They are more centrally concentrated than the
lower red giants of comparable brightness at the 97.9 \% level.
Their photometric characteristics imply that their majority
is not consistent with the collisional origin. We also reliably
isolated the asymptotic giant branch (AGB) clump and estimated the
parameter $\Delta V_{ZAHB}^{clump}$ = 1.01 $\pm$ 0.06, that is the
difference between the $V$-levels of the zero age HB and the clump.}
{}


\maketitle

\section{Introduction}
\label{introduc}

The southern Galactic globular cluster (GC) NGC 1261 ($\alpha_{2000}
= 03^{h}12^{m}15\fs3$ and $\delta_{2000} =
-55\degr13\arcmin01\arcsec$) remains one of the rarely studied GCs
up to now, in spite of its moderate central concentration ($c$ =
1.27), negligible or no reddening because of the relatively high
Galactic latitude ($b$ = $-52\fdg13$), and the somewhat brighter
absolute visual magnitude (M$_{V}$ = -7.81) compared to the most
probable value of this characteristic of Galactic GCs (Harris
\cite{harris96}). Although the history of CMD studies of this object
spans almost 40 years, it resulted so far in only six publications.
The first, scattered photographic CMD reaching the horizontal branch
(HB) of the cluster, was obtained by Alca\'ino \& Contreras
(\cite{alcontrer71}) and another one of a higher quality, by
Alca\'ino (\cite{alcaino79}). The first CCD photometry of NGC 1261,
aimed at estimating the cluster's age, was carried out twenty years
ago by Bolte \& Marleau (\cite{boltemar89}). Since then three
photometric studies were published in 1990s: Alca\'ino et al.
(\cite{alcainoetal92}) focused mainly on the age determination,
based on photometry in three bands; Ferraro et al.
(\cite{ferraroetal93}) studied the brighter sequences of the CMD,
and Zoccali et al. (\cite{zoccalietal}) obtained the luminosity and
mass functions of NGC 1261. Yet, Hubble Space Telescope photometry
in the cluster has been carried out by Piotto et al.
(\cite{piottoetal}), among 74 Galactic GCs.  Note than our short
overview is not concerned with publications dealing with variable
stars in NGC 1261 and other particular issues.

In this work, we study for the first time the stellar population of
NGC 1261 based on photometry that covers a wide cluster field
(approximately $14\arcmin \times 14\arcmin$) and was conducted in
several bands, including measurements in the $U$ bandpass, which is
rarely used in investigations of Galactic GCs. In particular, we
estimated the cluster metalicity by applying metalicity indicators
not previously used in NGC 1261, one of which is related to the
$U$-based CMDs. Also, signs of a probable inhomogeneity of the
stellar population in NGC 1261 were found with $U$ photometry in the
cluster.

\begin{figure*}
\centerline{
\includegraphics[clip=,angle=0,width=8 cm]{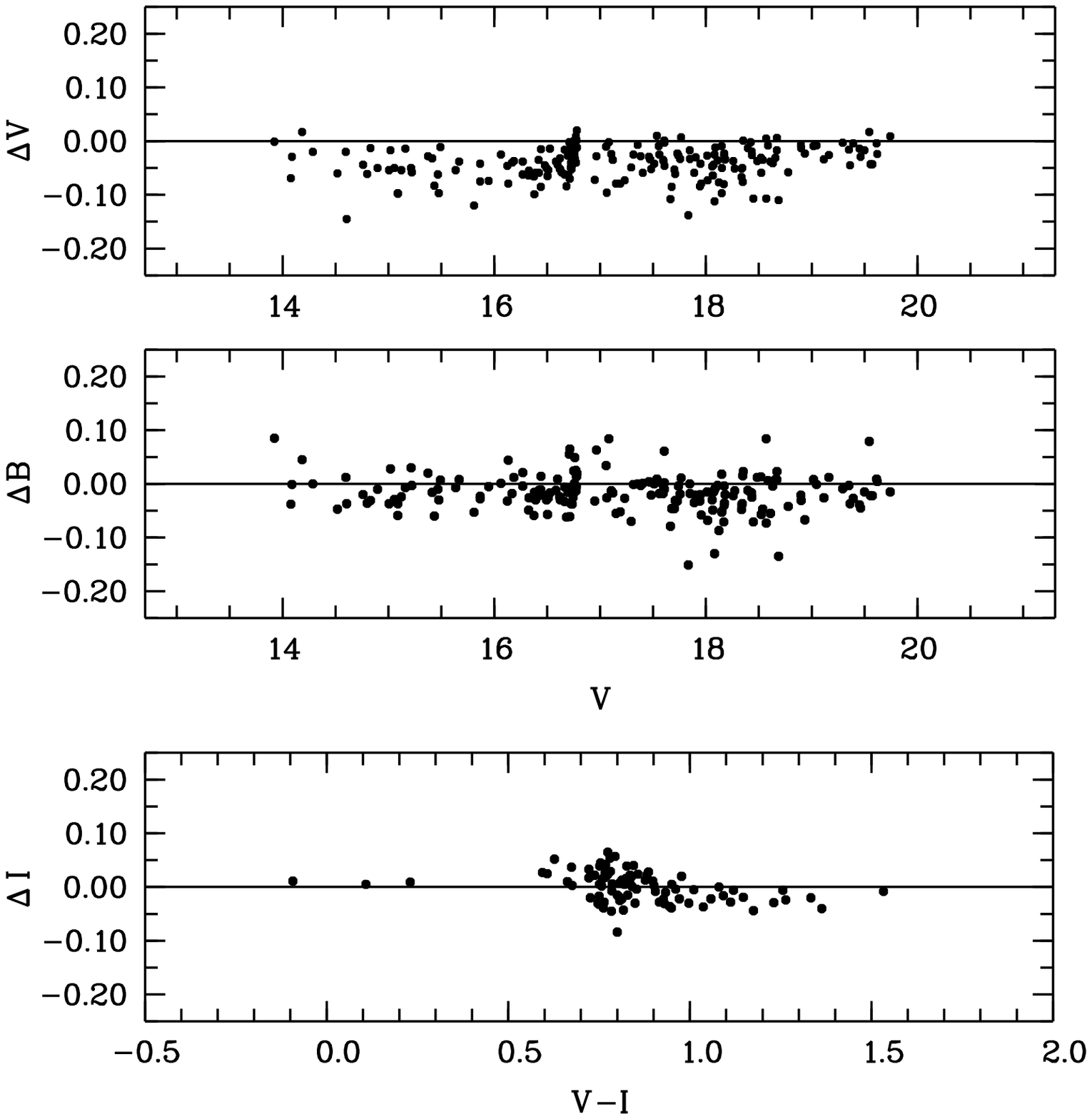}
\includegraphics[clip=,angle=0,width=8 cm]{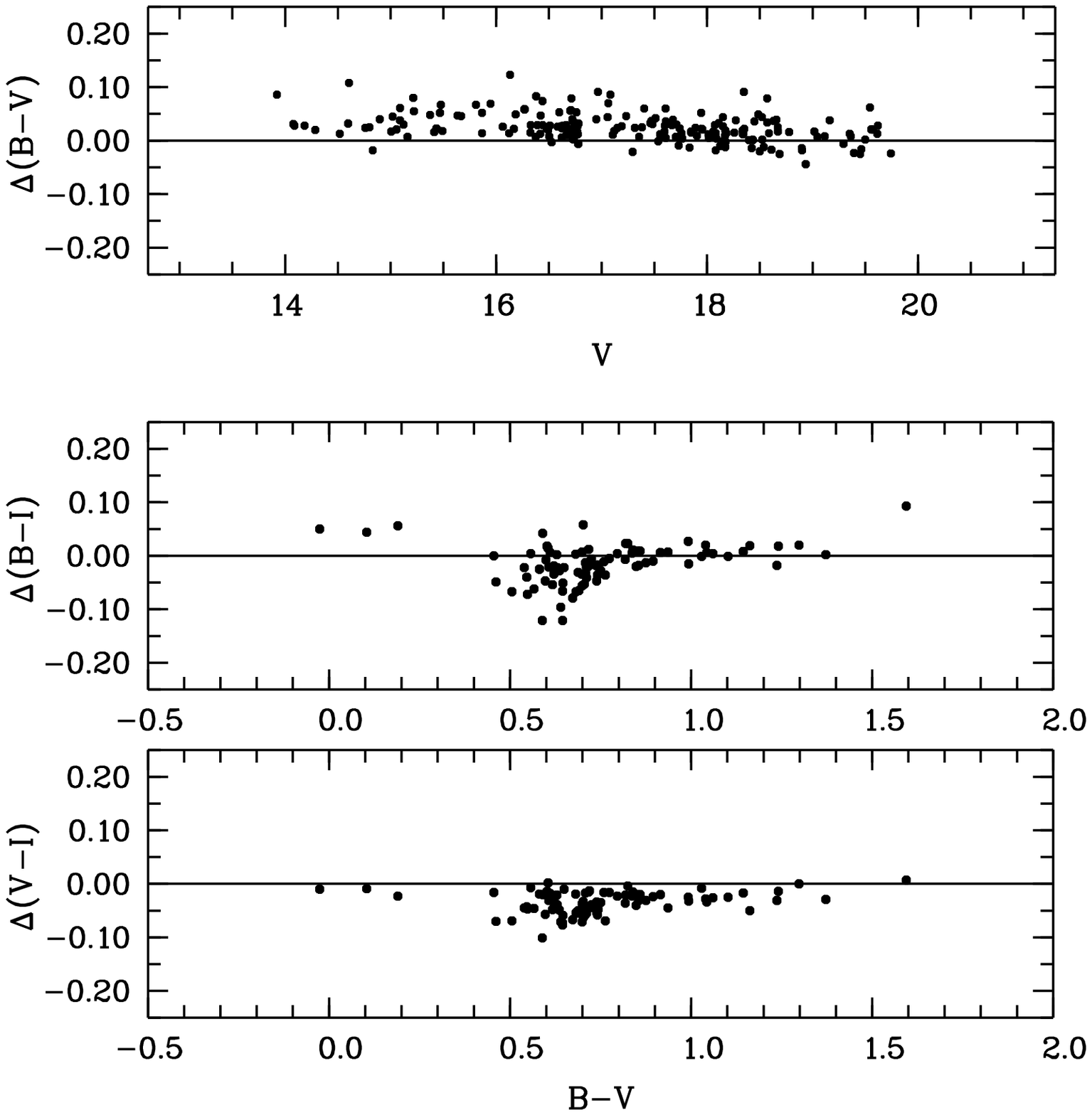}
} \caption{Comparison of our present photometry of NGC 1261 and the
photometry by Stetson (\cite{ste00}) of standard stars in the
cluster, taken from the Stetson Photometric Standard Fields; the
differences in magnitudes and colors are derived by our data minus
those of Stetson.} \label{comparphot}
\end{figure*}

\section{Observations and data reduction}
\label{dataredu}

The observations were made on five nights, 1997 December $27/28$,
$29/30$, $30/31$, and 1988 January $01/02$, $02/03$, with the 1.3 m
Warsaw telescope, Las Campanas Observatory, using a set of $UBVI$
filters and a $2048 \times 2048$ CCD camera with a gain $=3.8$ and a
readout noise of $5.5e^-$ rms. The array scale was $0\farcs417 {\rm
pixel}^{-1}$, giving a field of view of $14\arcmin \times
14\arcmin$. The center of the measured field of NGC 1261 was
approximately $10\arcsec$ to the west and $50\arcsec$ to the south
of the cluster center. Flat-field, bias, and dark frames were taken
twice per night, at the beginning and at the end. We took a total of
15 frames in $U$ (exposure time from 180 sec to 720 sec), 15 frames
in $B$ (60 sec to 360 sec), 15 frames in $V$ (40 sec to 240 sec),
and 15 frames in $I$ (30 sec to 180 sec). The observations were
gathered with the air mass varying between 1.146 and 1.698. The
average seeing on each night, estimated from the observations, was
in the range of approximately $1\farcs1$ to $1\farcs2$.

The reductions of CCD photometry were performed at the Isaac Newton
Institute and at the European Southern Observatory, ESO, Santiago,
Chile.  The stellar photometry was carried out separately for all
frames with the {\sc daophot/allstar} (Stetson \cite{ste87},
\cite{ste91}) and IRAF package\footnote{IRAF is distributed by the
National Optical Astronomy Observatory, which is operated by the the
Association of Universities for Research in Astronomy, Inc, under
cooperative agreement with the National Science Foundation.}. The
program stars were detected and measured by applying the usual
procedures. To obtain the PSF 20 to 30 stars in each frame, bright
but far from  saturation, were selected among those without
neighbors or defects within the PSF radius. We found that among
standard PSFs provided by DAOPHOT, the PENNY2 function enabled us to
handle the aberrations specific to individual frames most
effectively. The instrumental magnitudes and colors ($v$, $u-b$,
$b-v$, $v-i$) obtained for the measured stars were then transformed
to the standard system ($V$, $U-B$, $B-V$, $V-I$) using 25 standard
stars, of which 13 stars ($UBVI$) were from Alca\'ino \& Liller
(\cite{al84}), seven stars ($UBVI$) from Alca\'ino \& Liller
(\cite{al88}), and another five stars ($UBV$) from Alca\'ino \&
Contreras (\cite{alcontrer71}).

We used the "least squares" method to calculate a straight line that
best fitted the data for these standards stars. The equations used
in this study to bring instrumental magnitudes and colors to the
standard $UBVI$ photometric system are

$$V=v-0.01(\pm0.01)(b-v)+0.01(\pm0.03),  n = 25,  $$
$$V=v-0.03(\pm0.02)(v-i)+0.05(\pm0.02),  n = 15,   $$
$$U-B=1.22(\pm0.02)(u-b)-0.06(\pm0.01),  n = 11,  $$
$$B-V=1.00(\pm0.02)(b-v)+0.05(\pm0.01),  n = 25,  $$
$$V-I=0.92(\pm0.03)(v-i)+0.11(\pm0.02),   n = 15.  $$

The standard errors of the slope coefficients and constants are
given along with the number $n$ of standard stars used for the
respective transformations.

\begin{figure*}[t!!]
   \centering
   \includegraphics[angle=-90,width=18cm]{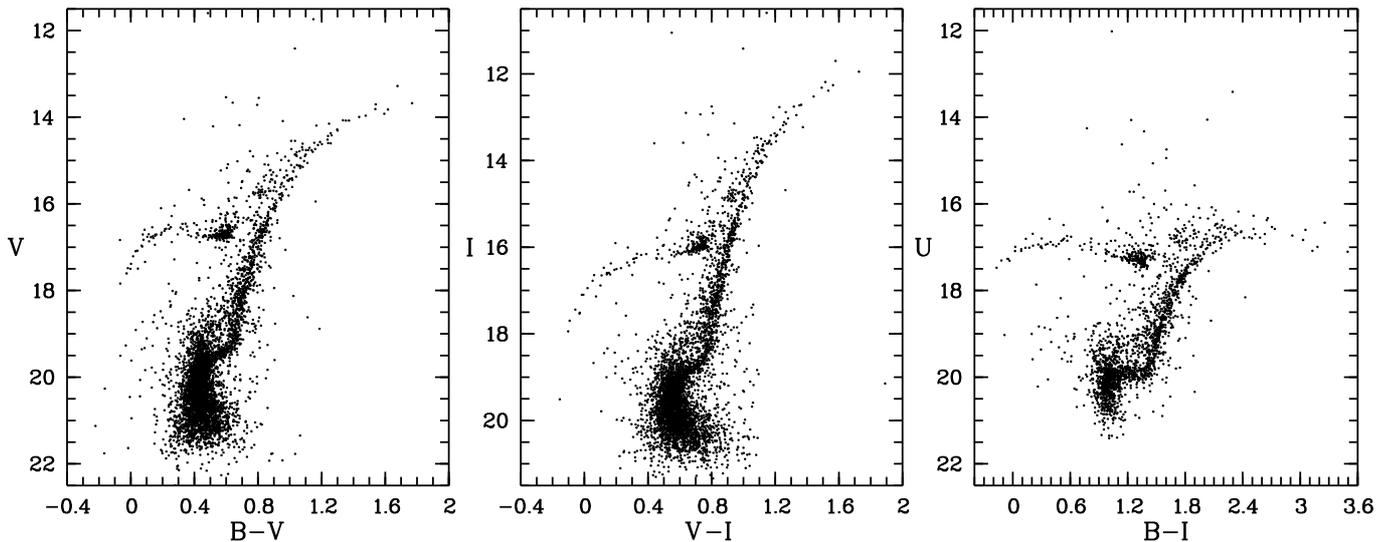}
         \caption{The $V$-$(B-V)$, $I$-$(V-I)$ and $U$-$(B-I)$ CMDs of the globular cluster
NGC 1261 based on $UBVI$photometry of 5\,481 stars in a
14$\arcmin$x14$\arcmin$ field approximately centered on the
cluster.}
         \label{cmd}
   \end{figure*}

In our preliminary list we retained only those stars that had at
least two measurements per night in each photometric band. We
included in our final list only those stars for which at least one
color-index was determined. A total of 5\,481 stars were measured
mainly in  $V$. Of these, 2157 stars have $U-B$ color, 5005 have
$B-V$, and 4841 have $V-I$\footnote{The results of our photometry
are available in electronic form}. For stars of the brighter
sequences, i.e. with $V < 19.0$, the r.m.s. errors are on average
$\pm0.013$ in $U$, $\pm0.015$ in  $B$, $\pm0.011$ in $V$, and
$\pm0.008$ in $I$.

We compared our photometry with the photometry of Stetson
(\cite{ste00}) of around 400 cluster stars in $BVI$, with $V <$ 20.0
mag. It was taken from the Stetson Photometric Standard Fields
available at
http://www3.cadc-ccda.hia-iha.nrc-cnrc.gc.ca/community/STETSON/stds/.
We were able to identify 196 stars in the field that were the same.
This comparison is shown in Fig.~\ref{comparphot}. It demonstrates
the very good agreement between the two photometries both in
magnitudes and colors, over the entire range. Apart from very small
or negligible differences in zero-points (within a few hundreds of
magnitude) of the photometries, we did not find any notable
systematic effect. The comparison also confirms the really good
accuracy of our photometry at least up to the limiting $V$-magnitude
(close to the cluster turnoff point) of the Stetson standard stars.
The mean differences between the two photometries in the sense of
this work minus that of Stetson (\cite{ste00}), and the number of
stars with $V <$ 20.0 utilized for the calculations are as follows:
$\delta V = -0.04 (\pm0.03)$, 196; $\delta B = -0.02 (\pm0.03)$,
196; $\delta (B-V) = 0.03 (\pm0.03)$, 196; $\delta (V-I) = -0.04
(\pm0.02)$, 88; $\delta (B-I) = -0.01 (\pm0.04)$, 88. The quoted
uncertainties are standard deviations.

We also compared our photometry with that by Ferraro et al.
(\cite{ferraroetal93}) and found 2494 stars in common. Again we did
not find any obvious systematic effect in color or in magnitudes,
especially for a sub-sample of the brighter stars of this sample
with $V < 19.0$. There are only small zero-point offsets: a positive
one, within 0.05 mag in the $V$ magnitude, and a negative in the
$(B-V)$ color, within a few hundreds of magnitude for this work
minus that of Ferraro et al.

\section{The color-magnitude diagrams}

\subsection{General comments}

In the panels of Fig.~\ref{cmd} we show the $V$-$(B-V)$,
$I$-$(V-I)$, and $U$-$(B-I)$ color-magnitude diagrams (CMDs)
constructed with different colors and magnitudes of our photometry
of all 5\,481 stars measured in the observed field of NGC 1261. Due
to crowding effects in its inner part, the overall appearance of the
CMDs is somewhat worsened by the more scattered sequences in the
crowded regions. Increasing the number of blended stars towards the
crowded central regions is thought to be (mainly?) responsible for
the appearance of numerous stars above the sub-giant branch (SGB)
and blueward of the red giant branch RGB, in the form of an apparent
sequence. The same feature is also seen in the CMD obtained by
Ferraro et al. (\cite{ferraroetal93}). In the subsequent analysis of
our photometry, we excluded the most crowded central region of NGC
1261.

Because of the fairly high Galactic latitude of the cluster, the
contamination of its CMDs by field stars is low. The field stars
become detectable only in CMDs of large outer parts of the observed
cluster field. In our subsequent analysis we reduced the effect of
the contamination on the reliability of the obtained results where
it was necessary with multi-color photometry.

\subsection{The sub-giant branch}
\label{sgb}

Both a visual inspection of the cluster CMDs and a more objective
analysis of our photometry showed not only the apparently thicker
sub-giant branch (SGB) in the $U$-based CMDs than in CMDs with
longer wavelength magnitudes with respect to the $U$, but also the
radial variation in the cluster from the mean level of the SGB.

\begin{figure}
  \centering
 \includegraphics[angle=-90,width=6cm]{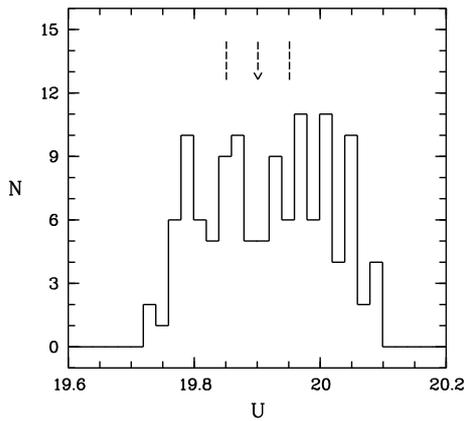}
      \caption{Distribution of the $U$ magnitude of a total sample of 122 SGB
stars isolated in a large field of NGC 1261, excluding its inner
part, as described in detail in the text. The arrow shows an
apparent minimum at $U =$ 19.90 mag, while the two dashed lines are
the boundaries of a magnitude range $\Delta U =$ 0.10, centered on
the minimum, with two neighboring magnitude ranges by $\Delta U =$
0.15 (see explanations in the text).}
         \label{sgbUmagdistr}
   \end{figure}

\begin{figure}
  \centering
 \includegraphics[angle=-90,width=7cm]{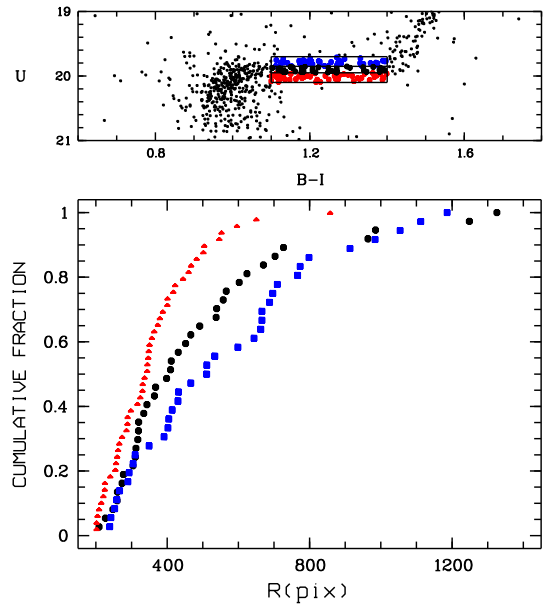}
      \caption{Upper panel shows three sub-samples of the SGB stars isolated
       in three magnitude ranges in the $U$-$(B-I)$ CMD of a less crowded
field located at radial distances larger than 200 pixels
($1\farcm36$) from the cluster center. Lower panel: a comparison of
the cumulative radial distributions of the three sub-samples of SGB
stars; red filled triangles, black filled circles, and blue filled
squares denote the sub-samples of the SGB stars with progressively
increasing luminosity in the $U$ passband. In both panels, the three
sub-samples are shown with the symbols of the same color.}
         \label{sgbdistrib}
   \end{figure}

To derive quantitative estimates and thereby to achieve more
objective conclusions, we proceeded in the following way. We first
obtained a sample of SGB stars. For this purpose the $U$-based
cluster CMD with the $(B-I)$ color-index was used, because it
provides the largest separation in a color between the turnoff point
(TO) and the lower RGB compared with other available color-indices.
To avoid the negative impacts of crowding effect on the obtained
results as much as possible, the central part of the cluster,
confined within $r = 1\farcm36$ (200 pixels), was excluded from the
analysis. Due to fainter magnitudes of SGB stars, the limiting
radius of the excluded region is larger than that applied for
selection of brighter stars belonging in particular to the HB or RGB
bump analyzed below. The SGB of NGC 1261 in the $U$-$(B-I)$ CMD is
virtually horizontal, which facilitates the selection of stars and
the subsequent analysis. A raw sample of candidate SBG stars was
selected in the color range $\Delta (B-I) =$ 0.3 ($1.1 < B-I < 1.4$)
and the magnitude range $\Delta U =$ 0.4 mag ($19.7 < U < 20.1$). To
clean it of possible contamination by some field stars and by stars
deviating from the TO region and the transitional region between the
SGB and lower RGB, the selected stars were plotted in CMDs with
magnitudes other than $U$. The most deviating stars were rejected.
The final cleaned sample contained 122 items. We found that the
distribution of the selected stars on the $U$ magnitude is
apparently rather bimodal than unimodal, with the minimum around $U
=$ 19.90, which is just at the middle point of the selection
magnitude range (Fig.~\ref{sgbUmagdistr}). Based on this
distribution, the obtained sample of the SGB stars was divided in
three sub-samples: (1) a sub-sample of 36 brightest SGB stars in the
magnitude range $\Delta U =$ 0.15 ($19.70 < U < 19.85$); (2) a
sub-sample of 49 faintest SGB stars in the magnitude range $\Delta U
=$ 0.15 ($19.95 < U < 20.10$); (3) a sub-sample of 37 SGB stars with
intermediate brightness in the $U$, falling around the minimum in
the magnitude range $\Delta U =$ 0.10 ($19.85 < U < 19.95$). The
selection boxes are drawn by the solid line in the $U$-$(B-I)$ CMD
in the upper panel of Fig.~\ref{sgbdistrib}, where the corresponding
selected stars of the three sub-samples with progressively
increasing brightness are denoted by red, black, and blue filled
circles, respectively.

The lower panel of Fig.~\ref{sgbdistrib} shows the cumulative radial
distributions of the three sub-samples of SGB stars. One can see
that the faintest sub-giants denoted by (red) filled triangles are
more centrally concentrated than their counterparts of intermediate
brightness in the $U$ (black filled circles), and obviously yet more
centrally concentrated than the brightest SGB stars (blue filled
squares). This apparent difference between the distributions is
supported by a quantitative estimate based on a Kolmogorov-Smirnov
(K-S) test: the difference between the distributions of the
brightest and faintest SGB stars is statistically significant at a
99.9\% confidence level. Note that the stars falling in the
brightest and faintest boxes are (almost) indistinguishable in the
$V$ and $I$ magnitudes.

This statistically significant dependence of the radial distribution
of SGB stars in NGC 1261 on their $U$ magnitude is strong evidence
of the inhomogeneity ("multiplicity") of the cluster stellar
population. This inhomogeneity of SGB stars resembles the split of
the SGB into two components revealed by Milone et al.
(\cite{milonetal08}) in the globular cluster NGC 1851 and the
different radial distribution of stars belonging to the brighter and
fainter components of the SGB, found by Zoccali et al.
(\cite{zoccalietal09}). According to the latter authors, the
brighter SGB stars are less centrally concentrated and extend to
much larger radial distance in NGC 1851. But the results of Milone
et al. (\cite{milonetal09}) do not support this finding though.

Unfortunately we are not able to draw any definite conclusion about
whether or not the revealed differences among the SGB stars are
discrete or continuous.

\subsection{The red giant branch}
\label{rgb}

The demonstrated evidence of the inhomogeneity of SGB stars is
supported by effects found in the RGB.

To analyze the RGB of NGC 1261, we first isolated the most probable
RGB stars. We minimized the contamination of the RGB in three steps
(1) stars belonging to both the asymptotic giant branch (AGB) and
red HB (RHB), (2) a possible few field stars that can appear among
the RGB stars in CMD with a given color-index, but are displaced
from the sequence on CMDs with other color-indices or/and on the
two-color diagrams or (3) stars showing a considerable deviation
from the sequence's fiducial line due to photometric error. For this
purpose we used the advantage of a multi-color photometry and
proceeded like this. In the $V$-$(B-I)$ CMD we fitted the mean locus
of the RGB with a polynomial applying corresponding commands in the
MIDAS system. We then linearized the RGB by subtracting for each
star the color of the mean locus at its luminosity level from the
star's color-index. We left only those stars that satisfied our
selection criterion: their deviations $\delta(B-I)$ from the mean
locus did not exceed $\pm$ 0.06 mag in the total luminosity range of
the RGB from its base to the tip. These conditional boundaries of
the RGB in the color-index separate the bulk of its stars from the
majority of stars belonging to both the AGB and the RHB and are
close to the mean errors in the colors along the RGB. It cannot be
excluded that some of stars really belonging to the RGB near its tip
were rejected or that contrarily some AGB stars merging with those
in the upper RGB near its tip were left. The isolated stars were
then plotted in CMDs with other color-indices to reject a few of the
stars deviating most from the RGB.

\begin{figure}
  \centering
 \includegraphics[angle=-90,width=7cm]{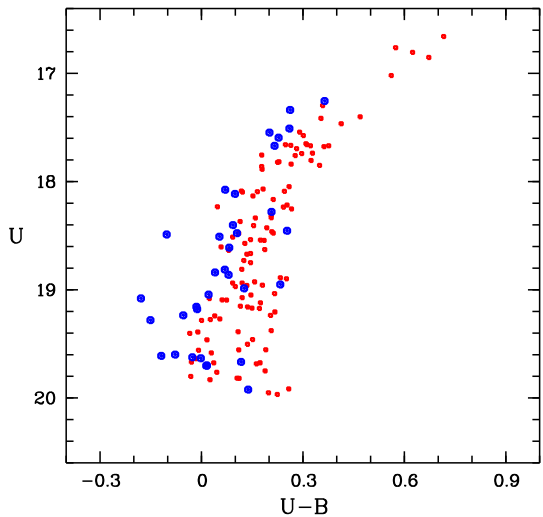}
   \caption{Comparison of the position in the $U$-$(U-B)$ CMD
of RGB stars from the inner (red dots, $1\farcm02 < r < 1\farcm70$)
and outermost (blue filled circles, $r > 4\farcm08$) regions of NGC
1261.}
         \label{rgbposit}
   \end{figure}

Based on the finally obtained sample of 448 RGB stars (which have
measurements at least in two passbands, $B$ and $V$), we were able
to develop and reinforce the result on the inhomogeneity of SGB
stars. We analyzed specially the expected relationship between the
photometric and spatial characteristics of RGB stars. We obtained
strong evidence of the inhomogeneity of the population of RGB stars
in NGC 1261 as a result. It is based on the dependence of their
location in the $U$-$(U-B)$ CMD on the radial distance from the
cluster center. Fig.~\ref{rgbposit} demonstrates this dependence
from a comparison of the position in this diagram of RGB stars from
the inner ($1\farcm02 < r < 1\farcm70$; 116 stars) and the outermost
(blue filled circles, $r > 4\farcm08$; 33 stars) regions of NGC
1261. It is evident that red giants at a larger radial distance from
the cluster center are on average bluer in the $(U-B)$ color (or/and
brighter in the $U$ magnitude). In this connection, we note that
Marino et al. (\cite{marinoetal08}) find a dichotomy in Na abundance
in a sample of 105 stars in the GC M4 and argue that it must be
associated with a CN bimodality. From photometry of the same stars
they also showed on the other hand that the CN-weak stars with a
simultaneously lower content of Na are systematically bluer, by
$\Delta (U-B) = 0.17$ in the $U$-$(U-B)$ CMD than their CN-strong
counterparts with a higher content of Na. This photometric effect
resembles that revealed by us in NGC 1261. It is quite difficult to
obtain a reliable estimate of the mean separation in the $(U-B)$
color between "bluest" and "reddest" RGB stars. It seems to be of
the same order of magnitude as the separation deduced by Marino et
al. (\cite{marinoetal08}) for the RGB of M4. However, for NGC 1261
we are inclined to adopt a more conservative estimate of $\Delta
(U-B) \sim 0.10$. Hence as a first approximation one can adopt the
same reason that led to the segregation of RGB stars in the $(U-B)$
in both GCs, NGC 1261 and M4. At the same time, we make the opposite
suggestion about a probable evolutionary association between the
discussed sub-populations of SGB/RGB stars and those belonging to
the BHB and RHB. Our suggestion is mainly based on the similarity
between the radial distributions in NGC 1261 of the sub-populations
of SGB/RGB stars on the one hand and the radial distributions of BHB
and RHB stars on the other. More details are given in
Sect.~\ref{hb}.

\begin{figure}
  \centering
 \includegraphics[angle=-90,width=7cm]{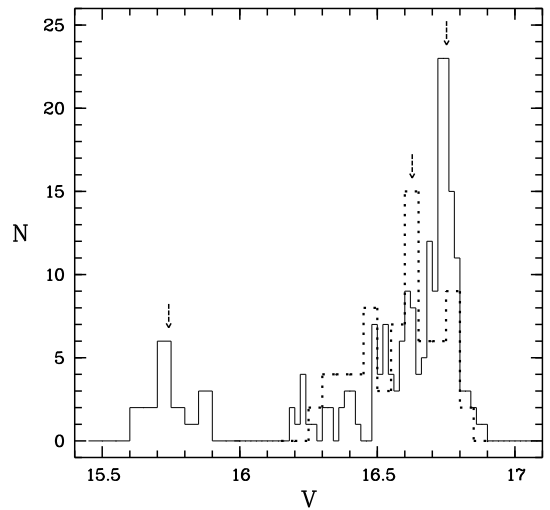}
      \caption{Distributions of the $V$ magnitude of stars belonging to the RGB bump
(dotted line), AGB clump (solid line), and the RHB (solid line at
fainter magnitude, with smaller bin size of the histogram); the
arrows show the adopted location of the two features and of the
ZAHB.}
         \label{bumclumhb}
   \end{figure}

With the obtained sample of RGB stars we constructed the RGB
luminosity function (LF) in the $V$ magnitude, identified the RGB
bump, and isolated a sample of 70 stars that fall in the bump region
(BR), i.e. in the magnitude range $\Delta V =$ 0.6 ($16.25 < V <
16.85$), just between the two nearest deepest minima of the LF on
both sides of the RGB bump. This fragment of the RGB LF is shown by
the dotted line in Fig.~\ref{bumclumhb}. The most probable location
of the bump was found at $V_{bump} = 16.63 \pm 0.03$. It is close
to, but is slightly brighter than $V_{bump} = 16.70 \pm 0.05$
deduced by Ferraro et al. (\cite{ferraroetal93}). This difference
seems to be primarily due to the larger sample size of the RGB bump
stars in our photometry, which therefore allowed us to estimate the
location of the RGB bump more confidently by relying on a smaller
bin of the LF.

\begin{figure}
  \centering
 \includegraphics[angle=-90,width=6cm]{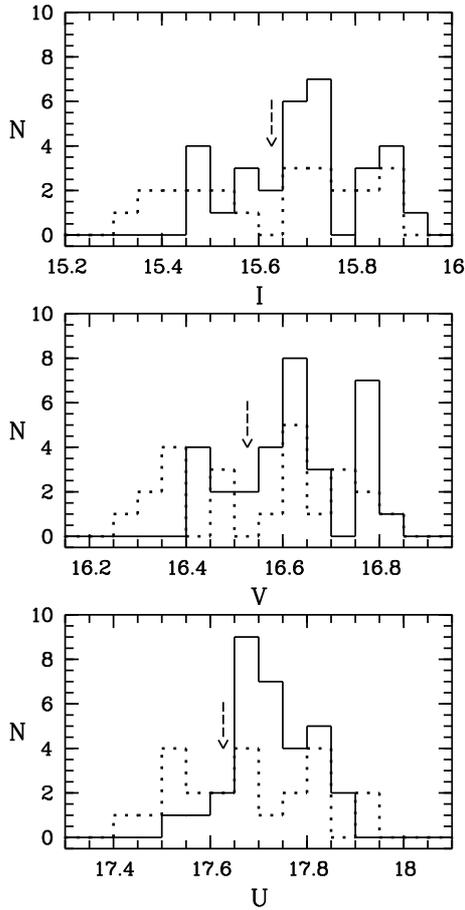}
   \caption{Upper, middle, and lower panels show the magnitude distribution
of the RGB bump stars in the $I$, $V$, and $U$ bands, respectively.
In each panel, the solid and dotted lines compare the distributions
in the inner ($0\farcm68 < r < 1\farcm70$; 31 stars) and in the
outer ($r > 1\farcm70$; 23 stars) regions respectively of the
cluster; the arrow indicates the adopted separation between the main
and brighter modes of the RGB bump in each plot.}
         \label{bumpdistrib}
   \end{figure}

By analyzing the BR LFs in different magnitudes, we found an
apparent radial variation in these LFs, which is particularly
obvious in the $U$ passband. The BR LFs are shown in
Fig.~\ref{bumpdistrib}, where the solid line is the BR LF of the
inner part ($0\farcm68 < r < 1\farcm70$; 31 stars) and the dotted
line is the BR LF of the outer part of NGC 1261 ($r > 1\farcm70$, or
250 pixels; 23 stars). To reduce the effect of the increasing
photometric error toward the cluster center on the results of our
analysis, the innermost region of the cluster containing sixteen
stars of the sample was excluded from our consideration. In the
panels of Fig.~\ref{bumpdistrib}, one can fix two kinds of
differences between the BR LFs in two parts of the cluster. First,
the main maximum that defines the position of the RGB bump dominates
in the inner region and almost disappears in the BR LF of the outer
cluster part, with the effect apparently more pronounced in the $U$
magnitude. Second, the relative number of stars $N_{brt}/N_{fnt}$ in
the brighter and fainter parts of the BR respectively, increases
with radial distance from the cluster center. In addition, the BR
LFs are systematically more extended toward higher luminosity, by
$\sim$ 0.1 to 0.15 mag, in the outer cluster regions. The adopted
boundary between the two parts of the BR in each magnitude ($I_{b}$,
$V_{b}$, and $U_{b}$) is marked by the arrow in the three panels of
the figure.

To calculate the ratio as carefully as possible, we minimized the
effect of the photometric scatter on the estimated ratio as much as
possible. For this purpose, we ascribed only those stars to the
brighter and fainter parts of the BR whose magnitudes are brighter
and fainter respectively, than the adopted boundary magnitudes
between the two parts of the BR, i.e. ($I < I_{b}$, $V < V_{b}$, $U
< U_{b}$ simultaneously) and ($I
> I_{b}$, $V > V_{b}$, $U > U_{b}$ simultaneously). Hence we
obtained $N_{brt}/N_{fnt} = 3/23 = 0.13 \pm0.08$ in the inner region
($0\farcm68 < r < 1\farcm70$) in contrast to $N_{brt}/N_{fnt} = 8/12
= 0.67 \pm 0.34$ in the outer part of the cluster ($r > 1\farcm70$).
The radial trend of the ratio is obvious. However, given the errors
the difference is marginal. We also estimated the level of
statistical significance of the difference between the BR LFs in two
parts of the cluster for all three passbands. The K-S test rejects
at 85.9\%, 85.9\%, and 92.4\% confidence levels the hypothesis that
the BR LFs of the inner and outer parts of NGC 1261 are the same
distributions in the $I$, $V$, and $U$ passbands, respectively.
Although the estimated confidence levels are relatively high,
especially in the $U$ magnitude, the difference between the BR LFs
in the two parts of NGC 1261 are not, strictly speaking,
statistically significant, because no confidence levels are greater
than 95\%. For this reason the discussed apparent radial variations
in the BR LFs may be considered as a suspected supplementary sign of
the inhomogeneity of RGB stars.

\subsection{The horizontal branch}
\label{hb}

Ferraro et al. (\cite{ferraroetal93}) found a probable radial trend
of the ratio of the number of BHB stars, $N_{BHB}$, to that of RHB,
$N_{RHB}$, in the sense that $N_{BHB}/N_{RHB}$ increases with
increasing radial distance from the cluster center. Based on our
photometry, which extends to a larger radial distance, we estimate
the ratio in two parts of the cluster, located at different mean
radial distance: $N_{BHB}/N_{RHB} = 10/77 = 0.13 \pm 0.05$ for
$0\farcm68 < r < 1\farcm70$ and $N_{BHB}/N_{RHB} = 13/42 = 0.31 \pm
0.10$ for $r
> 1\farcm70$. To avoid an artificial trend, we excluded the most
crowded innermost region of the cluster. Although we cannot yet
exclude a somewhat higher relative incompleteness of BHB stars in
the region located at shorter radial distance, we find the same
trend, i.e. RHB stars are presumably more centrally concentrated
than BHB ones. Formally, calculated separately in the outermost
region, the ratio becomes yet higher, but its statistical
significance is negligible due to the very low number of stars used.

We recall that both the SGB stars brighter in the $U$ magnitude and
the RGB stars bluer in the $(U-B)$ color are very probably less
centrally concentrated in NGC 1261 than their counterparts with
distinct photometric characteristics (i.e. the SGB stars fainter in
the $U$ and the RGB stars redder in the $(U-B)$, respectively). At
the same time, their radial distribution in the cluster resembles
just that of BHB stars. Moreover, the relative number of the
$U$-brighter SGB stars and (probably) that of the $(U-B)$-bluer RGB
stars is less than 1. This seems to generally agree with the
relative number of the BHB stars, i.e. the ratio $N_{BHB}/N_{RHB} <
1$. Given these characteristics, we assume that just the $U$-bright
SGB stars and the $(U-B)$-blue RGB stars are progenitors of of BHB
stars in the GC NGC 1261.  Note however that Marino et al.
(\cite{marinoetal08}) arrived at opposite conclusion about the RGB
progeny on the HB: they associated the Na-rich stars, the redder
stars in the $(U-B)$ color in the GC M4, with the BHB.

At the end we note that the $V$-level of the zero age HB (ZAHB) was
estimated to be at $V_{ZAHB} = 16.75 \pm 0.05$ (see
Fig.~\ref{bumclumhb}). Below we use this value to estimate some
useful parameters.

\subsection{The asymptotic giant branch clump}
\label{rgbbump}

The CMDs in Fig.~\ref{cmd} show that the number of stars tracing the
asymptotic giant branch (AGB) makes it possible to unambiguously
identify an important feature, the so-called AGB clump at the base
of the branch. It is clearly seen in all diagrams. In the
$U$-$(U-B)$ diagram though it is somewhat stretched along the color
axis.

The formation of the AGB clump as well as the well-known RGB bump is
caused by a slowing down of the rate of stellar evolution along the
given evolutionary sequence(s). For more details concerning the
nature of the clump and useful parameters deduced for it from the
CMD, as well as for more references related to the subject, we refer
in particular to Ferraro et al. (\cite{ferraroetal99}) and Sandquist
\& Bolte (\cite{sanqbolte}). In the present paper, we were able to
estimate one of the parameters with good accuracy, namely the
difference between the $V$-levels of the ZAHB and the AGB clump,
$\Delta V_{HB}^{clump}$. Because the AGB is poorly populated, the
estimates of the given parameter are available for the limited
number of GCs.

The AGB clump is located at $V_{clump}$ = 15.74 $\pm$ 0.03 (
Fig.~\ref{bumclumhb}). Taking into account that the ZAHB was
estimated to be at $V_{ZAHB} = 16.75 \pm 0.05$, we obtain the
parameter $\Delta V_{ZAHB}^{clump}$ = 1.01 $\pm$ 0.06.

\subsection{Blue stragglers}
\label{blst}

Relying on four-color photometry of the cluster we were able to
isolate blue straggler (BS) candidates more confidently than with
two-color photometry. This was achieved with the requirement of
simultaneous location of the candidates within the region of the
most probable concentration of BSs in several CMDs with different
color-indices, namely $(B-V)$, $(V-I)$, and $(B-I)$. Doing so, we
isolated 18 probable candidates. This is quite a poor population of
BSs. The small number of the found BS candidate can (at least
partially) be explained by the increased incompleteness of stars
with the same brightness in our photometry, especially in the
crowded central part of the cluster. The location of the isolated BS
candidates in the $U$-$(B-V)$ CMD is shown in Fig.~\ref{metalicity}.

\begin{figure}
  \centering
 \includegraphics[angle=-90,width=7cm]{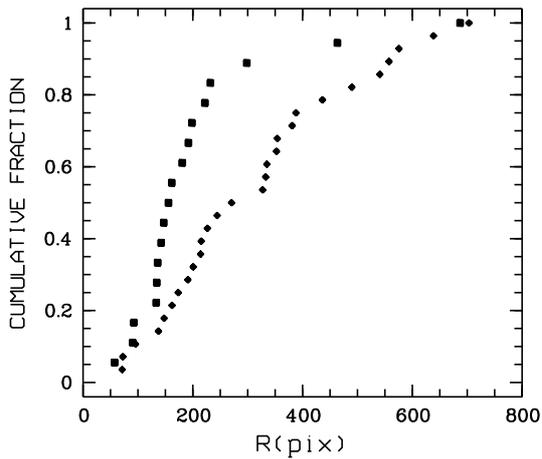}
\caption{Comparison of the cumulative radial distributions of blue
straggler candidates (filled squares) and lower RGB stars (filled
lozenges) falling in narrow magnitude range ($18.8 < V < 19.0$) that
is located at the middle level of the magnitude range spanned by the
blue straggler candidates ($17.8 < V < 20.0$).}
         \label{cumdistBS}
   \end{figure}

We compared the cumulative radial distribution of the isolated BS
candidates with that of a sample of 28 lower RGB stars in a narrow
magnitude range ($18.8 < V < 19.0$) that is located at the middle
level of the magnitude range spanned by the blue straggler
candidates ($17.8 < V < 20.0$). Fig.~\ref{cumdistBS} shows the
cumulative radial distribution of BS candidates and of the mentioned
RGB stars. One can see that the BS candidate are more centrally
concentrated than the RGB stars. This apparent obvious difference
between both distributions is confirmed by K-S test: the difference
is statistically significant at a 97.9\% confidence level.

The BS candidates appear in a high range of magnitude (more than 2
mag) in both $U$ and $V$, but they do not show any notable
segregation in the $U$ based two-color diagram, which otherwise
might imply distinct groups originating from different BS formation
mechanism. We found such two groups in M80 (Alca\'ino et al.
\cite{alcainoetal98}) by following the approach of Lauzeral et al.
(\cite{lauzaurcuop}), who in turn had applied the method of Bailyn
(\cite{bailyn}) to isolate BSs of probable collisional origin. We
conclude that among the BS candidates isolated by us in the observed
field, excluding the most crowded central region of NGC 1261 where
the completeness of faint stars is low and the photometry is
shallower, there is none whose origin is consistent with the
collisional mechanism.

The isolated BS candidates are marked by flags in the table with the
photometric data, which is available in electronic form.

\section{Metalicity}
\label{met}

To estimate the metalicity of NGC 1261, two methods were applied for
the first time for this GC.

\begin{figure}
  \centering
 \includegraphics[angle=-90,width=6.5cm]{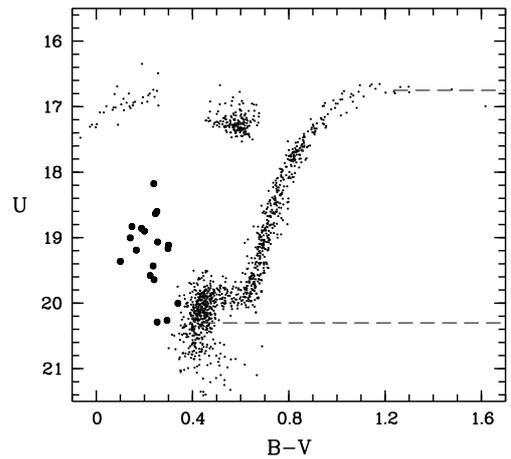}
   \caption{Composite, cleaned $U$-$(B-V)$ CMD of NGC1261 with
the most probable blue straggler candidates superimposed (filled
squires). the $U$-levels of the RGB inflection point and the TO
point are shown by the upper and lower horizontal dashed lines
respectively, aimed at demonstrating the metalicity-sensitive
parameter $\Delta U_{TO}^{RGB}$.}
         \label{metalicity}
   \end{figure}

We first estimated the metalicity of the cluster by relying on the
dependence of the position of the RGB bump on metalicity. This
dependence was empirically obtained and demonstrated for the first
time by Kravtsov (\cite{kravtsov89}). Here we use the analytically
expressed relation between metalicity and $\Delta V_{bump}^{ZAHB}$,
the magnitude difference between the bump, and the level of the
zero-age HB taken from Ferraro et al. (\cite{ferraroetal99}). As we
already noted, the most probable locations of the bump and the ZAHB
(see Fig.~\ref{bumclumhb}) were found to be at $V_{bump} = 16.63 \pm
0.03$ and $V_{ZAHB} = 16.75 \pm 0.05$, respectively. Hence $\Delta
V_{bump}^{ZAHB} = -0.12 \pm 0.06$ mag, which is converted to [Fe/H]$
= -1.41 \pm 0.10$ dex. Note that the same method applied to the data
deduced by Ferraro et al. (\cite{ferraroetal93}) gives a somewhat
higher metalicity by 0.2 dex. Indeed, they find the RGB bump
position at $V_{bump} = 16.70 \pm 0.05$ and the ZAHB level at the
same magnitude, $V_{ZAHB} = 16.70 \pm 0.04$. Hence one can calculate
$\Delta V_{bump}^{ZAHB} = 0.00 \pm 0.06$ mag, which translates to
[Fe/H]$ = -1.23$. The same level of the ZAHB and the RGB bump,
resulting in this metalicity value is different from our data
though: as is seen in Fig.~\ref{bumclumhb}, the bump and the
expected level of the ZAHB (and even somewhat brighter parts of the
HB) are clearly separated, with the bump somewhat brighter.

The above-obtained value of [Fe/H] agrees well with the metalicity
deduced from another metalicity indicator, the parameter $\Delta
U_{TO}^{RGB}$. In Kravtsov et al. (\cite{kravtsovetal}) we defined
this parameter as the difference in the $U$ magnitude between the TO
and the inflection point of the RGB in the $U$-$(B-V)$ CMD of GCs in
the metal-poor range at [Fe/H]$_{ZW} < -1.1$ dex in the scale of
Zinn \& West (\cite {zinnwest}). The color difference between the
two points is close to $\Delta (B-V) \approx$ 0.8 and almost
independent of the metalicity in the given metalicity range. We
found that the parameter $\Delta U_{TO}^{RGB}$ apparently does not
depend on variations in the response curve of the $U$ passband, at
least under the conditions of low reddening and the insignificant
red leak of the $U$ filters used. However, the parameter very
tightly correlates with GC metalicities. We also calibrated $\Delta
U_{TO}^{RGB}$ on the metalicity and deduced an analytical relation
between $\Delta U_{TO}^{RGB}$ and [Fe/H]$_{ZW}$.

We determined the parameter $\Delta U_{TO}^{RGB}$ using the
$U$-$(B-V)$ CMD. However, to estimate the $U$ magnitude of the TO
point as carefully as possible, we examined its apparent position in
the $U$-based CMDs with different color-indices and also compared it
with the $U$-level of those stars in the same CMDs, which fall in
the TO region in the $V$-based CMDs where the position of the TO
point can be estimated more reliably. We found the $U$-levels of the
RGB inflection point and of the TO point at $U_{RGB} = 16.75 \pm
0.13$ and $U_{TO} = 20.30 \pm 0.15$, respectively. They are shown by
two horizontal dashed lines in the $U$-$(B-V)$ diagram presented in
Fig.~\ref{metalicity}. Hence the parameter is close to $\Delta
U_{TO}^{RGB} = 3.55 \pm 0.20$ mag. Substituting this value in the
analytical relation between $\Delta U_{TO}^{RGB}$ and metalicity, we
obtained [Fe/H]$_{ZW} = -1.34 \pm 0.16$ dex. This value formally
coincides with that quoted in the catalog of Harris
(\cite{harris96}) and agrees well with the above-obtained quantity.
Hence the mean metalicity is [Fe/H]$_{ZW} = -1.38 \pm 0.14$ dex.

In the range of previously made estimates of the cluster's
metalicity (see, for example, compilations in Alca\'ino et al.
\cite{alcainoetal92} and Ferraro et al. \cite{ferraroetal93}) with
different methods and indicators, both our estimates (or their mean
value) fall near its metal-poor extreme. Therefore our estimates
disagree with the values of [Fe/H]$\sim -1$ dex close to the
metal-rich extreme of the cluster metalicity estimates, but agree
well with that adopted by Ferraro et al. (\cite{{ferraroetal93}}).

\section{Conclusions}

We obtained new $UBVI$ photometry of 5\,481 stars in a wide field,
approximately 14$\arcmin$x14$\arcmin$, of the southern, rarely
studied GC NGC 1261 below its TO point in all passbands. We
exploited this multi-color photometry especially in $U$ passband and
its advantages over two-color photometry to further study this GC.

We found two very probable indications and one suspected sign of the
inhomogeneity (multiplicity) of the stellar population in NGC 1261.
First, there is an obvious dependence of the radial distribution of
SGB stars in the cluster on their $U$ magnitude: brighter stars are
less centrally concentrated at the 99.9 \% confidence level and less
numerous than their fainter counterparts. Second, RGB stars exhibit
a systematically different location in the $U$-$(U-B)$ diagram at
different radial distance from the cluster center: the proportion of
stars bluer in the $(U-B)$ (or/and brighter in the $U$) increases
towards the cluster outskirts. Additionally, the distribution of
stars on magnitude, particularly in the $U$ band, in the RGB bump
region is suspected to change radially in the cluster. We conclude
that both the SGB stars brighter in $U$ and the RGB stars bluer in
the $(U-B)$ may be associated with the blue BHB, because they are
less numerous and less concentrated to the cluster center as
compared to their counterparts in the respective evolutionary
sequences.

We estimated the metalicity of NGC 1261 by applying metalicity
indicators, which were previously not used on this cluster. One of
them was recently proposed by us. It is related to deep, $U$-based
CMDs of GCs that simultaneously reach the TO point and trace the
upper RGB well with a sufficient amount of stars. The given
condition was accomplished by our photometry. We obtained
[Fe/H]$_{ZW} = -1.34 \pm 0.16$ dex. We also found [Fe/H]$_{ZW} =
-1.41 \pm 0.10$ by exploiting the dependence of the position of the
RGB bump on metalicity. Both estimates agree very well with each
other. The resulting mean value of the cluster's metalicity,
[Fe/H]$_{ZW} = -1.38 \pm 0.14$ dex, is at the metal-poor extreme of
the range of previously made estimates on the same metalicity scale
obtained by different methods and indicators.

We isolated 18 BS candidates, a quite small population of these
stars. They are found to be more centrally concentrated than the
lower red giants of a comparable brightness. This difference is
statistically significant at the 97.9 \% confidence level according
to the K-S test. Their position in the $U$-based two-color diagram
indicates that the formation of their majority is not due to the
collisional mechanism.

\begin{acknowledgements}
This research used the facilities of the Canadian Astronomy Data
Centre operated by the National Research Council of Canada with the
support of the Canadian Space Agency. FA is very grateful to Profs.
Drs. Felix Mirabel and Massimo Tarenghi, former and actual
representatives of ESO in Chile, respectively, for kindly providing
living and computing facilities at the ESO office in Santiago, where
the observations were reduced. We thank the anonymous referee for
useful comments that improved the manuscript.

\end{acknowledgements}


\begin{thebibliography}{}

\bibitem[1979]{alcaino79} Alca\'ino, G. 1979, \aaps, 38, 61

\bibitem[1971]{alcontrer71} Alca\'ino, G., \&  Contreras, C. 1971, \aap, 11, 14

\bibitem[1984]{al84} Alca\'ino, G., \&  Liller, W. 1984, \aj, 89, 1712

\bibitem[1988]{al88} Alca\'ino, G., \&  Liller, W. 1988, \aj, 96, 139

\bibitem[1992]{alcainoetal92} Alca\'ino, G., Liller, W., Alvarado,
F., \& Wenderoth, E. 1992, \aj, 104, 1850

\bibitem[1998]{alcainoetal98} Alca\'ino, G., Liller, W., Alvarado,
F., et al. 1998, \aj, 116, 2415

\bibitem[1989]{boltemar89} Bolte, M., \& Marleau, F. 1989, \aap, 101, 1088

\bibitem[1992]{bailyn} Bailyn, C. D. 1992, \apj, 392, 519

\bibitem[1993]{ferraroetal93} Ferraro, F. R., Clementini, G., Fusi Pecci, F., Vitiello, E.,
\& Buonanno, R. 1993, \mnras, 264, 273

\bibitem[1999]{ferraroetal99} Ferraro, F. R., Messineo, M., Fusi Pecci, F., et al. 1999,
\aj, 118, 1738

\bibitem[1996]{harris96} Harris, W. E. 1996, \aj, 112, 1487

\bibitem[1989]{kravtsov89} Kravtsov, V. V. 1989, Soviet. Astron. Lett., 15, 356

\bibitem[2007]{kravtsovetal} Kravtsov, V., Alca\'ino, G., Marconi, G., \& Alvarado, F.
2007, \aap, 469, 529

\bibitem[1993]{lauzaurcuop} Lauzeral, C., Auriere, M., \& Coupinot, C.
1993, \aap, 274, 214

\bibitem[2008]{marinoetal08} Marino, A. F., Villanova, S., Piotto, G. et al. 2008, \aap, 490,
625

\bibitem[2008]{milonetal08} Milone, A.P., Bedin, L.R., Piotto, G., et al. 2008, \apj, 673,
241

\bibitem[2009]{milonetal09} Milone, A.P., Stetson, P.B., Piotto, G., et al. 2009, \aap, 503,
755

\bibitem[2002]{piottoetal} Piotto, G., King, I. R., Djorgovski, S. G., et al. 2002, \aap,
391, 945

\bibitem[2004]{sanqbolte} Sandquist, E. L., \& Bolte M. 2004, \apj, 611,323

\bibitem[1987]{ste87} Stetson, P.B. 1987, \pasp, 99, 191

\bibitem[1991]{ste91} Stetson, P.B. 1991, DAOPHOT II Users  Manual

\bibitem[2000]{ste00} Stetson, P.B. 2000, \pasp, 112, 925

\bibitem[1984]{zinnwest} Zinn, R., \& West M. J. 1984, \apjs, 55, 45

\bibitem[1998]{zoccalietal} Zoccali, M., Piotto, G., Zaggia, S.R., \& Capaccioli, M. 1998,
\aap, 331, 541

\bibitem[2009]{zoccalietal09} Zoccali, M., Pancino, E., Catelan, M., et al. 2009,
\apj, 697L, 22

\end{thebibliography}
\end{document}